\documentclass[prl,
twocolumn,
preprintnumbers,showpacs,
superscriptaddress]{revtex4-1}
\usepackage{graphicx}% Include figure files
\usepackage{bm}% bold math
\usepackage{amsmath}
\usepackage{amssymb}
\usepackage[colorlinks,pdfstartview=FitH]{hyperref}
\usepackage{comment}
\hypersetup{linkcolor=blue,citecolor=blue,filecolor=black,urlcolor=blue}

\newcommand\xiJB{\xi_{JB}}
\newcommand\xiJo{\xi_{J\omega}}
\newcommand\xiTB{\xi_{TB}}
\newcommand\xiTo{\xi_{T\omega}}
\newcommand\xiSB{\xi_{SB}}
\newcommand\xiSo{\xi_{S\omega}}

\newcommand\KB{K_{B}}
\newcommand\Ko{K_{\omega}}
\newcommand\aB{\alpha_{B}}
\newcommand\aom{\alpha_{\omega}}
\newcommand\DeT{\tilde T}
\newcommand\DeJ{\tilde J}
\newcommand\DeS{\tilde S}
\newcommand\Force{{\cal F}}
\renewcommand\section[1]{\emph{#1}\ ---}

\begin{document}

\title{The no-drag frame for anomalous chiral fluid}
\author{Mikhail A.~Stephanov}
\affiliation{Department of Physics, University of Illinois, Chicago,
  Illinois 60607}
\author{Ho-Ung Yee}
\affiliation{Department of Physics, University of Illinois, Chicago,
  Illinois 60607}
\affiliation{RIKEN-BNL Research Center, Brookhaven National Laboratory
Upton, New York
11973-5000}

\preprint{RBRC-1146}
\begin{abstract}

  We show that for an anomalous fluid carrying dissipationless chiral magnetic and/or vortical currents there is a frame in which a stationary obstacle experiences no drag, but energy and charge currents do not vanish, resembling superfluidity. However, unlike ordinary superfluid flow, the anomalous chiral currents do transport entropy in this frame. We show that the second law of thermodynamics completely determines the amounts of these anomalous non-dissipative currents in the ``no-drag frame'' as polynomials in temperature and chemical potential with known anomaly coefficients. These general results are illustrated and confirmed by a calculation in the chiral kinetic theory and quark-gluon plasma at high temperature.

\end{abstract}

\maketitle

\section{Introduction}
\label{sec:intro}
The collective dynamics of a chiral (parity-violating) medium
associated with quantum anomalies has become a subject of much
attention recently. In particular, currents along the direction of an
external magnetic field (chiral magnetic effect, or CME) discussed
hypothetically earlier \cite{PhysRevD.22.3080} have been recently proposed in
Ref.~\cite{Fukushima:2008xe,Fukushima:2010vw} as a possible
explanation of the charge dependent correlations observed in heavy-ion
collisions and of negative magnetoresistance in a
Dirac-semimetal~\cite{Son:2012bg,Li:2014bha}. Currents in the
direction of rotation axis (chiral vortical effect, or CVE) also
discussed in astrophysical context before~\cite{PhysRevD.20.1807} have
been (re)discovered in strong-coupling gauge-gravity calculations
\cite{Erdmenger:2008rm,Banerjee:2008th}.  The generality of these
effects and their connection to chiral anomaly have been demonstrated
in Ref.~\cite{Son:2009tf,Neiman:2010zi} by applying the constraint of the second law
of thermodynamics to the hydrodynamic equations for the anomalous chiral fluid.

One of the manifestations of the anomalous nature of the CME and CVE
currents is that these currents are dissipationless and do not lead to
entropy production, in contrast, e.g., to the ordinary Ohmic current
driven by electric field, but similar to the persistent superfluid currents. 
 We wish to gain further
understanding of the non-dissipative nature of the anomalous transport.

How can one distinguish the anomalous CME/CVE currents
from the uniform (shearless) inertial motion of the fluid as a whole
 in the same direction,
which also does not generate entropy and carries energy and charge? To
do this one needs to determine the ``rest frame'' of the ``normal''
component of the flow. This situation appears to be similar to the
Landau's two-fluid picture of the superfluid~\cite{LandauFluid}. The
superfluid flow is also non-dissipative and carries energy (mass) and charge.
Since it carries no entropy, one can define the rest frame of the
normal component as the frame where the entropy flow vanishes. It is
tempting to use the same criterion in the case of the anomalous CME/CVE flows. We shall show that, in general, this would not be correct,
i.e., the CME/CVE currents {\em do} carry entropy.

We propose that a natural way to define the ``rest frame'' of the
normal component is to insert an impurity, or an obstacle, obstructing
the flow as it is done, e.g., in Ref.~\cite{Rajagopal:2015roa} for a
gauge theory plasma with a heavy quark. In general, the flow will
exert force on the obstacle, and if the obstacle is free to move it
will accelerate until it reaches a certain velocity at which the drag,
and thus acceleration, vanishes. One can say that the impurity will
then be carried by the flow and it defines the ``no-drag frame'' -- a
natural (physically meaningful) rest frame of the fluid.

We shall present a very general argument based on the
second law of thermodynamics which allows one to determine such a
no-drag frame, and thus the magnitude of the energy, charge and
entropy currents in it. Of course, the magnitude of the drag force
experienced by an impurity depends on the properties of the
impurity itself and its interaction with the medium. However, the velocity of
the no-drag frame is completely determined by the second law of
thermodynamics and is insensitive to the details of
impurity-medium interactions, thus representing the intrinsic
property of the fluid itself.

If we apply this logic to a normal fluid in equilibrium, the no-drag
frame, of course, will be the Landau frame, and the flows of energy,
charge and entropy vanish in it. For a superfluid, the no-drag frame
is the frame where the normal component is at rest, the energy and
charge is carried by the superfluid component, and there is no entropy
current. Using the second law of thermodynamics we shall show that the
anomalous CME/CVE currents not only carry 
energy and charge, but also entropy in the no-drag frame
\footnote{In agreement with the results of
  Ref.\cite{Rajagopal:2015roa} obtained in a holographic model with
  CME/CVE currents, which found nonzero flows of energy and
charge in the no-drag frame, while the entropy flow
vanished only when there was no mixed gauge-gravity anomaly.}.

\section{Anomalous hydrodynamics with drag}
Let us consider a fluid flowing past a fixed point-like obstacle, or impurity,
for example, an infinitely heavy quark, or possibly a lattice of such
impurities, or a porous solid medium. The hydrodynamic equations then contain
an additional term due to the momentum
transfer between the impurities and the fluid:
\begin{equation}
  \label{eq:dT}
  \partial_\mu T^{\mu\nu} = F^{\nu\lambda}J_\lambda - \Force^\nu
\end{equation}
The local 4-momentum transfer from the fluid to the impurities, i.e.,
the drag force, per unit volume, $\Force^\nu$, depends on the
hydrodynamic variables and the 4-velocity of the impurity. We shall show
that this dependence is constrained by the second law of
thermodynamics up to an overall non-negative coefficient.

To simplify the analysis we shall consider a single-flavor anomalous
fluid obeying
\begin{equation}
  \label{eq:dJ}
  \partial_\mu J^\mu = C E\cdot B
\end{equation}
where $E^\mu=F^{\mu\nu}u_\nu$ and $B^\mu=(1/2)\varepsilon^{\mu\nu\alpha\beta}u_\nu
F_{\alpha\beta} = \widetilde F^{\mu\nu}u_\nu$ as in Ref.\cite{Son:2009tf}. 
%The generalization to multiple conserved currents is straightforward.
The constitutive equations are given by
\begin{multline}
  \label{eq:Tmunu}
  T^{\mu\nu} \equiv w u^\mu u^\nu + p g^{\mu\nu} +
  \DeT^{\mu\nu};\\ % \qquad 
  \DeT^{\mu\nu}=(\xiTo\omega^\mu + \xiTB B^\mu) u^\nu + (\mu\leftrightarrow\nu),
\end{multline}
where $w=\varepsilon+p$, and
\begin{equation}
  \label{eq:Jmu}
  J^\mu \equiv nu^\mu +\DeJ^\mu;\qquad 
\DeJ^\mu= \xiJo\omega^\mu + \xiJB B^\mu,
\end{equation}
where $\omega^\mu =
(1/2)\varepsilon^{\mu\nu\alpha\beta}u_\nu\partial_\alpha u_\beta$ and
$\xi$'s are the anomalous transport (CME and CVE) coefficients.
We do not write the usual dissipative terms due to viscosity and
conductivity/diffusion because they will contribute only to an
increase of the entropy. The corresponding second-law constraints
on the viscosity and conductivity terms are well known and their
inclusion will not affect our results. We can also
think that we are considering an equilibrium state of the fluid where
all dissipative terms have already vanished, except, possibly, for the drag
force on a test impurity.

The entropy current is given by
\begin{equation}
  \label{eq:ds}
  S^\mu = s u^\mu + \DeS^\mu;\qquad 
\DeS^\mu= \xiSo\omega^\mu + \xiSB B^\mu.
\end{equation}
where $s$ is the entropy density. 

The 4-vector $u^\mu$ defines the local reference frame of the fluid
(the frame in which $u^\mu=(1,\bm 0)$). The well-known freedom of
choice of this frame is used to optimize the form of the
equations. E.g., we can choose it to be the frame in which $T^{0i}=0$
(Landau) or $J^i=0$ (Eckart) or $S^i=0$ (entropy frame). These three
choices coincide for a normal fluid flow {\em in equilibrium}, as we
discussed above, but for the anomalous transport (CME or CVE) these
choices are different.

The freedom of choosing the local reference frame allows us,
starting from any choice, to redefine the velocity by, e.g., 
\begin{equation}\label{eq:u-ualpha}
u^\mu\to u^\mu+\aom \omega^\mu + \aB B^\mu,
\end{equation}
where $\aom$ and $\aB$ are
arbitrary coefficients. Then the anomalous transport coefficients 
in Eqs.~(\ref{eq:Tmunu}),~(\ref{eq:Jmu}) and (\ref{eq:ds}) would change
accordingly:
\begin{subequations}   \label{eq:coeff}
  \begin{align}
    & \xiTo \to \xiTo - w\aom;\qquad \xiTB \to \xiTB - w\aB;\\
    & \xiJo \to \xiJo - n\aom;\qquad \xiJB \to \xiJB - n\aB;\\
    & \xiSo \to \xiSo - s\aom;\qquad \xiSB \to \xiSB - s\aB;
  \end{align}
\end{subequations}

One can, and we shall, use this freedom to go to a frame which
is most suitable for a given purpose --  in our case, the frame where drag vanishes.

% Thus, by choosing $\alpha_\omega$ and $\alpha_B$ we can make (almost) any
% linear combination of $\xi_{X\omega}$ and that of $\xi_{XB}$, where
% $X=T,J,S$, vanish.  We shall see that a particular linear
% combination in each case stands out, and that will help us identify
% the natural rest frame of the fluid.

\section{The second law and the drag}
Combining equations~(\ref{eq:dT})--(\ref{eq:ds}) with the first law 
 $d\varepsilon = Tds + \mu dn$ we can calculate the divergence of
 the entropy current and use the second law as a constraint~\cite{Son:2009tf,Neiman:2010zi} to establish
 the relations (invariant under~(\ref{eq:coeff})) between the coefficients:
 \begin{align}
%   \begin{equation}
     \label{eq:1} %{\cal O}^{(1)}_\mu \equiv
     Td\xiSo+\mu d\xiJo - d\xiTo&=\Ko\,dp/w;\\
     % \\+\Ko(u\cdot \partial) u^\mu ;
%   \end{equation}
%   \begin{equation} 
     \label{eq:2}
     % {\cal O}^{(2)}_\mu \equiv
     Td\xiSB +  \mu d\xiJB - d\xiTB &= \KB\,dp/w %\\+ \KB (u\cdot \partial) u^\mu
     ;\\
%   \end{equation}
%   \begin{equation}
     \label{eq:3} %{\cal O}^{(3)}\equiv
     \xiJo-2 (T\xiSB+\mu\xiJB-\xiTB)&=-\Ko\,n/w;\\
%   \end{equation}
%\begin{equation}
  \label{eq:4} %{\cal O}^{(4)}\equiv 
\xiJB-\mu C &=-\KB\,n/w,
%\end{equation}
\end{align}
where we defined the following linear combinations:
\begin{equation} \label{eq:KKB}
\Ko \equiv
2T\xiSo+2\mu\xiJo-3\xiTo
; ~%\qquad\mbox{and}\quad
K_B \equiv
T\xiSB+\mu\xiJB-2\xiTB\,.
\end{equation}
With that, the heat production rate is given by:
\begin{equation}
  \label{eq:d.s}
  T(\partial\cdot S) =  \left(u - 
\frac{\Ko\omega + \KB B}{w}\right)\cdot\Force
\end{equation}
Note that under transformations in Eqs.~(\ref{eq:coeff})
\begin{equation}
  \label{eq:Kalpha}
  \Ko\to \Ko+ w \aom;\qquad \KB\to\KB + w\aB,
\end{equation}
which is also clear from Eqs.~(\ref{eq:u-ualpha}),~(\ref{eq:d.s}) and
the fact that the entropy production should not depend on our choice
of local reference frame $u^\mu$. 

Although the discussion can be
continued using arbitrary frame, we find it most convenient to fix the
frame now by conditions
\footnote{In fact, had we made this choice even earlier, while deriving
  Eq.~(\ref{eq:d.s}), we would not have needed to use equations of motion to
  transform $(u\cdot\partial)u^\nu$, which simplifies the derivation
as well as the results compared to
Refs.\cite{Son:2009tf,Neiman:2010zi}, as observed in Ref.\cite{Loganayagam:2011mu}.} 
\begin{equation}
  \label{eq:K0}
  \Ko=\KB=0.
\end{equation}

With this choice, the only remaining nonzero term
in the r.h.s.\ of Eq.(\ref{eq:d.s}) is $u\cdot \Force$. The requirement that
it be non-negative for an arbitrary fluid flow and heavy quark
4-velocity $U$ fixes vector $\Force$ up to an arbitrary, but non-negative,
coefficient, $\lambda_\Force\ge0$:
\begin{equation}
  \label{eq:flambda}
  \Force^\mu %= \lambda_f P_U^{\mu\nu} u_\nu 
= \lambda_\Force (u^\mu +
  U^\mu(u\cdot U))
\end{equation}
Both terms in Eq.~(\ref{eq:flambda}) are needed because fluid cannot do
work on a static impurity and thus $\Force^0$ must vanish in the frame
defined by $U$, i.e., $U\cdot \Force=0$. One can also arrive at this
condition by considering the 4-momentum of a heavy quark
$P^\mu=MU^\mu$ and requiring that the transfer of the 4-momentum
$\Force^\mu\sim dP^\mu/d\tau$ does not violate the mass-shell
condition $P\cdot P=M^2$.

The second law of thermodynamics
$ T(\partial\cdot S)=u\cdot \Force =
\lambda_\Force ((  u\cdot U)^2-1)\ge 0$ only constrains the {\em sign} of
$\lambda_\Force$, whose magnitude could be a (local) function of hydrodynamic variables
$\varepsilon$, $n$ as well as $ u\cdot U$ and
coordinates.

Eq.~(\ref{eq:flambda}) implies that $\Force=0$ when $U= u$, i.e.,
if the heavy quark (or impurity) is at rest in the reference frame
we chose. In other words, the frame where coefficients obey Eq.~(\ref{eq:K0})  
is the no-drag fame. When both velocities are small $U\approx (1,\bm V)$ and $
u\approx(1, {\bm v})$, the drag force is proportional to the relative
velocity $\bm \Force \approx - \lambda_\Force (\bm V - {\bm v})$, as one would
expect, and the heat production rate is $ u\cdot \Force \approx\lambda_\Force(\bm
V - {\bm v})^2$.

\section{Anomalous coefficients in the no-drag frame}
\label{sec:anom-flows}
In order to determine the anomalous flows of energy, charge and
entropy in the no-drag frame we need to determine the six coefficients
$\xi$ in this frame. There are six equations: two differential
equations~(\ref{eq:1}),~(\ref{eq:2}) and four algebraic
Eqs.~(\ref{eq:3}),~(\ref{eq:4}) and
(\ref{eq:KKB}) together with (\ref{eq:K0}). 
They can be solved easily, e.g., using
algebraic equations to express all coefficients in terms of $\xiSo$
and $\xiSB$ and then integrating Eq.~(\ref{eq:2}) which %  becomes
% \begin{equation}
%   \label{eq:DxiSB}
%   d(\xiSB/T)=0,
% \end{equation}
% and
gives
\begin{equation}
  \label{eq:XB}
  \xiSB=X_BT
\end{equation}
with an arbitrary constant $X_B$ and after that Eq.~(\ref{eq:1}) which % becomes
% \begin{equation}
%   \label{eq:DxiSo}
%   d(\xiSo/T^2) - 2X_B d(\mu/T)=0,
% \end{equation}
% and 
gives
\begin{equation}
  \label{eq:xiSo}
  \xiSo = 2X_B\mu T + X_\omega T^2
\end{equation}
with another arbitrary constant $X_\omega$.

Substituting Eqs.~(\ref{eq:XB}) and~(\ref{eq:xiSo}) back into the
algebraic equations we determine
remaining four transport coefficients $\xi$:
\begin{align}
\label{eq:xiJB}&  \xiJB = C\mu\\ 
\label{eq:xiJo}&  \xiJo = C\mu^2 + X_B T^2\phantom{\frac12}\\
\label{eq:xiTB}&  \xiTB  = \frac12 C\mu^2 + \frac12 X_BT^2\\
\label{eq:xiTo}&  \xiTo 
= \frac23 C\mu^3 + 2X_B\,\mu T^2 + \frac23X_\omega T^3
\end{align}

It is remarkable that in the no-drag frame anomalous transport
coefficients are {\em polynomials} in $T$ and $\mu$ \footnote{ We
  checked that these results are in agreement with the analysis
  performed in Ref.\cite{Neiman:2010zi} in Landau frame, where the
  derivation and the resulting form of the coefficients is more
  complicated (not polynomial). In terms of Ref.\cite{Neiman:2010zi}
  $X_B=2\beta$ and $X_\omega=3\gamma$.}. The polynomial coefficients
have been also found in calculations using Kubo formulas
(see~\cite{Landsteiner:2012kd} and refs. therein), in the calculations
in Ref.~\cite{Loganayagam:2011mu}, where the ansatz equivalent to
(\ref{eq:K0}) was used to bypass equations of motion, in kinetic
theory calculations~\cite{Loganayagam:2012pz} and in holographic
calculations in Ref.~\cite{Chapman:2012my}, where the frame was
defined by horizon normal 4-vector. Despite the
results suggesting a special nature of the frame, the
simple physical significance of it -- being the no-drag frame -- has not been
realized till now. 

It has been also suspected that the special frame may be
characterized by vanishing entropy flow
e.g.,~\cite{Megias:2013xla,Rajagopal:2015roa}, 
in particular, based on the superfluid analogy~\cite{Kalaydzhyan:2012ut}.
This is not true in general, as we shall now discuss.

\section{Entropy flow}
\label{sec:entropy-flow}
It is remarkable, but not unexpected, that, even though the fluid
carries energy flow, a static heavy quark
(impurity or obstacle) experiences no drag. This reflects
dissipationless, persistent nature of the anomalous currents, similar
to the superfluid currents.

For comparison, relativistic superfluid hydrodynamics is described by
  constitutive equations with 
  \begin{equation}
    \label{eq:TJS-sf}
    \DeT^{\mu\nu}=v^2\psi^\mu\psi^\nu,\
  \DeJ^\mu = -v^2 \psi^\mu,\ %\mbox{and ({\em N.B.})}\ 
 \DeS^\mu=0,
  \end{equation}
 where
  $\psi^\mu = \partial^\mu\phi + A^\mu$ and $\phi$ is the Goldstone
  field (phase) obeying Josephson
  equation {$u\cdot\psi = \mu$} \cite{Son:2000ht}.  Using $d\varepsilon = Tds +
  \mu dn + v^2 \psi\cdot d\psi$ one can again show that $T\partial\cdot S =
  u\cdot \Force$, i.e., superfluid flow does not contribute to drag. 

Superfluid flow does not transport entropy (\ref{eq:TJS-sf}).
  On the other hand, the anomalous entropy flow in the no-drag frame
  is proportional to the coefficient $X_B$ in Eq.~(\ref{eq:XB}) which
  is related~\cite{Landsteiner:2011cp,Jensen:2012kj} to the mixed
  gauge-gravity anomaly given by the trace of the charge generator. In
   cases when that coefficient is zero, e.g., considered in
  Ref.~\cite{Rajagopal:2015roa}, the entropy current is
  absent. However, this
  property does not hold more generally, as equations~(\ref{eq:XB})
  and~(\ref{eq:xiSo}) show.
% ~\footnote{ The entropy flow due to CVE was
    % calculated in a holographic model in
    % Ref.\cite{Chapman:2012my}.}.

We can verify these general results and better understand the physics
involved by using examples where the hydrodynamic behavior is
derivable from a microscopic description. We shall consider two such
examples: CVE in Lorentz invariant chiral kinetic theory with
collisions~\cite{Chen:2015gta} and CME in a chiral gauge plasma at
high temperature.

\section{Examples}
The Lorentz invariant chiral kinetic equation is given by
\begin{equation}
  \label{eq:dj}
  \partial\cdot j = {\cal C}[f]
\end{equation}
where $j^\mu$ is the covariant phase-space particle number current and ${\cal
  C}$ is the collision rate for a given distribution function $f$ (see
Ref.~\cite{Chen:2015gta} for details).
A uniformly rotating equilibrium
solution to this equation can be written as $f=(e^g + 1)^{-1}$ where~\cite{Chen:2014cla,Chen:2015gta}
\begin{equation}
  \label{eq:g=beta-mu-S-nobar}
  g = \beta u\cdot p + \frac12 \Omega_{\mu\nu}S^{\mu\nu}
  - \beta\mu q.
\end{equation}
The property which ensures the detailed balance and vanishing
of ${\cal C}[f]$ is that $g$ is a linear combination of quantities
{\em conserved} in each collision: 4-momentum $p^\mu$, angular momentum $S^{\mu\nu}$ and particle number (charge) $q$, where $\beta u_\mu$, $\Omega_{\mu\nu}$
and $\beta\mu$ are the coefficients~\cite{Chen:2015gta,Chen:2014cla}.

% correspond to inverse temperature, 4-velocity, vorticity and chemical potential,
% and $ \Omega_{\mu\nu} = \partial_\mu (\beta u_\nu)={\rm const}$.

Let us now show that for the solution given by
Eq.~(\ref{eq:g=beta-mu-S-nobar}) drag would be absent in the frame given by
$u^\mu$. To describe an impurity we need to add another term into
the kinetic equation describing collisions of the particles with the impurity:
\begin{equation}
  \label{eq:Cdrag}
 \partial\cdot j = {\cal C}+  {\cal C}_U;\qquad
  {\cal C}_U = \int_{AB} C_{AB} 
\end{equation}
\begin{equation}
  \label{eq:Cab}
  C_{AB} = W_{B\to A} - W_{A\to B}
\end{equation}
\begin{equation}
  \label{eq:WAB}
  W_{A\to B}[\bar f] = |M_2(E,\theta)|^2
\,  (2\pi)\,\delta(p_A\cdot U-p_B\cdot U)\,
%\left[
\,%\times
\bar f_A(1-\bar f_B) ,
%\right]
\end{equation}
where the matrix element is a function of two independent Lorentz invariants:
the energy $E=p_A\cdot U=p_B\cdot U$,  and the scattering angle
$\theta$, or $p_A\cdot p_B=E^2(1-\cos\theta)$ in the frame $U$. The
distribution function $\bar f$ is also evaluated in frame $U$.
The collisions in ${\cal C}_U$ do {\em not}
conserve 3-momentum, unlike those in ${\cal C}$. However, collisions are elastic and the energy {\em
  is} conserved in the frame $U$ where the impurity is at rest. 
% The distribution function $\bar f$ in Eq.~(\ref{eq:WAB}) is
% evaluated in the frame $\bar n = U$ (the center of mass of the
% collision is the rest frame of the infinitely heavy quark).  As in
% the case of $2\to2$ scattering, the collision kernel is
% nonlocal~\cite{Chen:2015gta}, which is essential to conserve angular
% momentum (and insure Lorentz invariance).

The drag force is given by the rate of momentum transfer from the
colliding particles to the impurity:
\begin{equation}
  \label{eq:fCfrag}
  F^\mu = \int_{AB} C_{AB}\, (p_A - p_B)^\mu 
\end{equation}
The energy conservation in Eq.~(\ref{eq:WAB}) ensures that \mbox{$U\cdot F=0$}. However, even for an equilibrium solution in
Eq.~(\ref{eq:g=beta-mu-S-nobar}) with generic $u^\mu$ the 3-momentum
transfer need not vanish -- the impurity will experience a drag,
as expected. Correspondingly, the distribution function obtained from
Eq.~(\ref{eq:g=beta-mu-S-nobar}) will not satisfy the detailed balance
condition $W_{B\to A} = W_{A\to B}$ because such $g$ is
not a linear combination of {\em conserved} quantities -- the
3-momentum in frame $U$ is not conserved in Eq.~(\ref{eq:WAB}). Thus
the solution to the kinetic equation (\ref{eq:dj}) given by 
Eq.~(\ref{eq:g=beta-mu-S-nobar}) will not solve the kinetic equation~(\ref{eq:Cdrag}). In other
words, the impurity will disturb the flow. 

However, when $u=U$, the component of momentum appearing in
Eq.~(\ref{eq:g=beta-mu-S-nobar}) $u\cdot p = U\cdot p$ (the energy in
frame $U$) {\em is} conserved according to Eq.~(\ref{eq:WAB}). Thus
detailed balance will be satisfied when $U=u$, i.e., $C_{AB}=0$.  This
then ensures that ${\cal C}_U=0$, i.e., the distribution in
Eq.~(\ref{eq:g=beta-mu-S-nobar}) with $u=U$ is still an equilibrium
solution, and that the drag force in Eq.~(\ref{eq:fCfrag}) vanishes
even though particles do scatter off impurity.

The CVE transport coefficients for the distribution in
Eq.~(\ref{eq:g=beta-mu-S-nobar}) have been calculated in Ref.\cite{Chen:2015gta}:
\begin{align}
  \label{eq:xis}
&  \xiJo = \frac{\mu^2}{4\pi^2} + \frac{ T^2}{12}\,;\\
&  \xiTo = \frac{\mu^3}{6\pi^2} + \frac{\mu T^2}{6}\,;\\
&  \xiSo = \frac{\mu T}{6}\,.
\end{align}
In agreement with Eqs.~(\ref{eq:xiJo}),~(\ref{eq:xiTo})
and~(\ref{eq:xiSo}) they are polynomials in $\mu$ and $T$, with
$C=1/(4\pi^2)$, $X_B=1/12$ and $X_\omega=0$.

As another example, we can consider CME in a non-abelian chiral
quark-gluon plasma at high temperature. It is known that anomalous
flows in the frame used to define canonical thermal density matrix
$e^{-\beta \hat H}$ agree with Eqs.~(\ref{eq:xis}) in non-interacting
limit~\cite{Landsteiner:2012kd}, and we expect the same even with
interactions.  To check the value of the drag, we can take the 1-gluon
scattering rate with momentum transfer $\bm q$ in this frame given
by~\cite{Pisarski:1993rf}
\begin{equation}
{\cal R}(\bm q)=\frac{d\Gamma}{ d^3 \bm q}
=\frac{\alpha_s T}{ 2\pi^2}C_R\lim_{q^0\to 0}
\frac{\rho_L(q^0,\bm q)}{ q^0}\,,
\end{equation}
where $C_R$ is the color Casimir and $\rho_L(q)=-2\,{\rm Im\,}G_R^{00}(q)$ is the longitudinal
gluon spectral density. Because $G^{00}_R$ is real in coordinate
space, the Fourier transform satisfies  $\rho_L(q^0,\bm
q)=-\rho_L(-q^0,-\bm q)$.
Since also $\rho_L(q^0,\bm q)\to q^0 f(\bm q)$ for $q^0\to 0$, we
must have
${\cal R}(\bm q)={\cal R}(-\bm q)$, and the drag force on the
heavy quark vanishes
\begin{equation}
\bm F=\int_{\bm q}\,\bm q\, {\cal R}(\bm q)=0
\end{equation}
independently of the external magnetic field.

\section{Conclusions}
\label{sec:conclusions}
We observed that the second law of thermodynamics constrains the
dependence of the drag force on the velocity (up to an overall
positive coefficient whose value is determined by the microscopic
details of the interaction between the obstacle and the fluid). In
equilibrium, the drag vanishes when the obstacle is at rest in a
certain frame associated with the fluid -- the no-drag
frame. Conversely, for a fluid flowing through a pipe, the drag
forces on the walls will dissipate the normal flows until they vanish
in the frame where the walls are at rest.  For a normal fluid in
equilibrium all conserved currents (energy, charge, entropy) vanish in
that rest frame.

However, as we have shown, for a fluid carrying anomalous CME or CVE
currents, the no-drag frame is characterized by certain {\em non-zero}
values of the anomalous currents proportional to the magnetic field or
vorticity with coefficients given by
Eqs.~(\ref{eq:XB})--(\ref{eq:xiTo}).

It is useful to invoke the Landau's two-fluid picture of superfluidity
\cite{LandauFluid}.  In this well-known picture, the normal component
of the flow exerts drag and generates entropy moving past an obstacle,
while the superfluid component, as the name implies, flows with no
drag. The no-drag frame is the frame where the normal flow
vanishes. The equilibrium superfluid flows of energy and charge,
however, do not vanish in that frame
(\ref{eq:TJS-sf}). Similarly, the anomalous CME and CVE flows of
energy and charge do not vanish in the no-drag frame. In this sense,
anomalous flows are also ``superfluid''.

This analogy breaks down when we consider the entropy flow. The
superfluid flow (proper) carries no entropy (\ref{eq:TJS-sf}). In contrast,
both CME and CVE anomalous currents do, in general, carry entropy (as well as
energy and charge), according to Eqs.~(\ref{eq:ds}),~(\ref{eq:XB}),~(\ref{eq:xiSo}).

The absence of the drag associated with anomalous chiral transport may
have phenomenological
implications~\cite{Rajagopal:2015roa}. It would be also interesting to
explore physical applications (e.g., in a thermomechanical effect) of the novel and unusual fact that persistent,
non-dissipative anomalous currents can carry entropy.

We thank  R.~Pisarski, K.~Rajagopal, A.~Sadofyev  and Y.~Yin for discussions. This work is
supported by the DOE grant No.\ DE-FG0201ER41195.

\bibliography{Drag}

\end{document}